\documentclass[aps,floatfix,twocolumn,a4paper,noshowpacs, nofootinbib,superscriptaddress,10pt]{revtex4}
\usepackage{graphicx,float}\usepackage{graphicx,float}
\usepackage[all]{xy}
\usepackage{amsmath,upgreek,bigints}
\usepackage{amssymb}
\usepackage{color}
\usepackage{epsfig,bm}		
\usepackage{graphicx,epstopdf}
\usepackage{subfigure}
\usepackage{pdfpages}
\usepackage[colorlinks,hyperindex]{hyperref}
 
  \newcommand{\clt}{\textcolor{black}}
    \newcommand{\cat}{\textcolor{black}}

        \newcommand{\svkps}{strange axial-vector kaon  resonances\;}
            \newcommand{\svka}{strange axial-vector kaons\;}

\definecolor{green1}{RGB}{0,128,0} 
\hypersetup{backref=true,pagebackref=true}
\hypersetup{%
  colorlinks = true,
  linkcolor  = blue,
  citecolor = cyan,
}
\usepackage{bookmark,textgreek}
\usepackage{hyperref,color,xcolor}
\hypersetup{hidelinks,hyperindex=true,colorlinks=true,breaklinks=true,urlcolor= blue}
\hypersetup{%
  colorlinks = true,
  linkcolor  = blue
}\usepackage{amssymb}
\newsavebox{\foobox}

\usepackage{graphicx,float,tikz}
\usepackage[all]{xy}
\newcommand\ringring[1]{%
  {
   \mathop{\kern0pt #1}\limits^{
     \vbox to-1.85ex{
       \kern-2ex 
       \hbox to 0pt{\hss\normalfont\kern.1em \r{}\kern-.45em \r{}\hss}%
       \vss 
     }
   }
  }
}
\newcommand\orcidgk{{\href{https://orcid.org/0000-0002-0785-2826
}{\orcidicon}}}
\newcommand\orcidroldao{{\href{https://orcid.org/0000-0003-3978-532X}{\orcidicon}}}
\newcommand\orcidwayne{{\href{https://orcid.org/0000-0002-7701-0421}{\orcidicon}}}
\newcommand{\orcidicon}{%
	\begin{tikzpicture}
	\draw[lime, fill=lime] (0,0)
		circle [radius=0.16]
		node[white] {{\fontfamily{qag}\selectfont \tiny ID}};
	\draw[white, fill=white] (-0.0625,0.095)
		circle [radius=0.007];
	\end{tikzpicture}	\hspace{-2mm}
}
\newcommand{\bpartial}{\mathop{\partial\kern -4pt\raisebox{.8pt}{$|$}}}

\newcommand{\bes}{\begin{subequations}}
\newcommand{\ees}{\end{subequations}}
\def\beq{\begin{eqnarray}}

\def\eeq{\end{eqnarray}}
\def\be{\begin{equation}}
\def\ee{\end{equation}}

\begin{document}

\title{Configurational information measure of mesonic states in 4-flavor  AdS/QCD}

\author{G. Karapetyan\orcidgk\!\!}
\email{gayane.karapetyan@ufabc.edu.br}
\affiliation{Federal University of ABC, Center of Mathematics, Santo Andr\'e, 09580-210, Brazil}

\author{W. de Paula\orcidwayne\!\!}
\email{wayne@ita.br}
\affiliation{Instituto Tecnol\'ogico de Aeron\'autica,  DCTA, S\~ao Jos\'e dos Campos, 12228-900, Brazil}

\author{R. da Rocha\orcidroldao\!\!}
\email{roldao.rocha@ufabc.edu.br}
\affiliation{Federal University of ABC, Center of Mathematics, Santo Andr\'e, 09580-210, Brazil}

\begin{abstract}
Strange axial-vector kaons, $K_1$, and $f_1$ meson  resonances are investigated in the 4-flavor AdS/QCD model. Their underlying differential configurational entropy is computed and the mass spectra of higher-excited resonances, in both these mesonic families, are achieved and discussed. This technique merges the 4-flavor AdS/QCD and experimental data regarding the mass spectrum of $K_1$ and $f_1$ meson resonances that have been already detected and reported in the Particle Data Group, also bringing forth a route to explore physical features of the next generation of resonances in the $K_1$ and $f_1$ meson families.

 \end{abstract}
\maketitle

\section{Introduction}

The differential configurational entropy (DCE) is a configurational information measure based on Shannon's information entropy, appraising the information entropy of any continuous physical system, also governing the correlation function  among the waves modes constituting it and measuring the order in the system  \cite{Gleiser:2018kbq,Gleiser:2018jpd}.  
Ref. \cite{Bernardini:2016hvx}  employed the DCE to scrutinize mesons in AdS/QCD, followed by tachyonic extensions \cite{Barbosa-Cendejas:2018mng}. The DCE has been addressing significant features of hadronic matter in the context of holographic QCD. The DCE can provide a verisimilar description of several features of confined hadrons with great accuracy. In particular, the mass spectrum  of hadronic resonances has been scrutinized using the tools of the DCE \cite{Bernardini:2018uuy,Braga:2017fsb,Karapetyan:2018oye,Karapetyan:2018yhm,Braga:2018fyc,Braga:2020myi,daRocha:2021imz}. The DCE apparatus unriddled  several  open problems regarding some meson families, as charmonium and the bottomonium, hybrid mesons, tetraquarks, quarkonium-like exotica,  tensor mesons, baryonic matter, hadrocharmonium, hadronic molecules,   glueball fields, and odderons    \cite{Colangelo:2018mrt,Ferreira:2020iry,Braga:2020hhs,Karapetyan:2019fst,Karapetyan:2021vyh,Ma:2018wtw,Braga:2021fey,Braga:2021zyi,Karapetyan:2021crv,Karapetyan:2017edu}, matching experimental data provided by the Particle Data Group (PDG) \cite{pdg}.  The DCE has been addressing other important features in the context of QFT and string theory in Refs. \cite{Mvondo-She:2023xel,Braga:2019jqg,Cruz:2019kwh,Lee:2019tod,Bazeia:2018uyg,Bazeia:2021stz,Lee:2018zmp,Chinaglia:2017dxf,Correa:2016utv,Koliogiannis:2022uim,Lima:2022yzk}.

AdS/QCD comprises an effective paradigm to investigate meson resonances \cite{Colangelo:2008us}. Holographic QCD takes in the non-perturbative regime of QCD,  with mesons being described by gauge fields of a Yang--Mills theory with flavor  \cite{Csaki,Karch:2006pv,Branz:2010ub,Bianchi:2010cy}. Among the bottom-up approaches, the  soft wall AdS/QCD requires a dilaton field coupled to gravity and can well describe Regge trajectories \cite{Karch:2006pv}. 
\cat{Knowing the spectrum of hadrons yields valuable information about the non-perturbative aspects of QCD, as well as the inner structure of hadrons. 
The analysis of the Regge trajectory of hadrons consists of one of the most efficient 
approaches for investigating the hadronic spectrum.  The Regge trajectories for mesons constituted of light-flavor quarks are linear with good accuracy, with a universal slope $\alpha\approx 0.85$ GeV${}^2$, which 
arises in models of quark confinement. 
When the strange quarks, as well as the bottom and charm quarks, are taken into account, the Regge  trajectory is not linear any longer \cite{Afonin:2014nya}.  Another reason for the nonlinearity of Regge  trajectories for heavy-flavor quarks is the screening of the quark-antiquark potential due to the additional $q\bar{q}$ pair production \cite{Gershtein:2006ng}.  The nonlinearity of Regge trajectories sets in when mesons with heavier flavor are taken into account, being more pronounced and noticeable for heavier masses of the constituent quarks. To implement non-linear Regge trajectories, a deformation of the quadratic dilaton  was employed and addressed in Ref. \cite{MartinContreras:2020cyg}, since Ref. \cite{Karch:2006pv} showed that a quadratic dilaton field correction in the AdS${}_5$ background yields a linear Regge behavior $M_n^2\propto n$, also avoiding the
ambiguities in the choice of boundary conditions at the IR wall. When only light-flavor quarks are regarded, the constraints imposed by chiral symmetry in the limit of
massless quarks yield  the quadratic dilaton. In particular,  the Regge behavior plays a prominent role in studying the hadronic spectrum, also  working as a probe to the inner structure of mesons, providing direct hints of the type of their quark constituents.  } Complementarily, it is possible to introduce confinement (Wilson loop area law criteria) dynamically, by solving the coupled 5D Einstein-dilaton equations \cite{Gursoy:2007cb,dePaula:2008fp,Li:2013oda,He:2013qq}.   Bottom-up AdS/QCD  complies with accurately fitting the QCD phenomenology as the asymptotic boundary field theory of gravity coupled to a gauge theory in the AdS bulk \cite{dePaula:2009za,Ballon-Bayona:2017sxa,Ballon-Bayona:2021ibm,MartinContreras:2022lxl,Afonin:2022hbb,MartinContreras:2020cyg,Afonin:2014nya}. 
Experimental data yields a well-established model to describe strange axial-vector $K_1$ mesons with positive parity, whose lighter family members have a similar quark composition. The quark model comprises two nonets,  ${}^3P_1$ and the ${}^1P_1$, of axial-vector meson states with  $J^P = 1^+$. 
The strange axial-vector
meson states in each nonet can mix to create mass eigenstates  $K_1(1270)$ and $K_1(1400)$, which thus have either an $s\bar{d}$ or a $d\bar{s}$ constitution. On the other hand, the $f_1$ meson family has $I^G(J^{PC})=0^+(1^{++})$ quantum numbers.

In this work, the DCE of the $K_1$ \svkps and the $f_1$ meson family will be computed and analyzed in the 4-flavor AdS/QCD apparatus, which takes into account the $u$, $d$, $s$, and $c$ quarks. 
Following Ref. \cite{Erlich:2005qh}, where it was presented a holographic chiral symmetry-breaking realization within the hard-wall model, Ref. \cite{Ballon-Bayona:2017bwk} proposed an extension to  four flavors, which encompasses chiral and flavor symmetry breaking. The procedure consists of implementing an expansion of the 5D fields into Kaluza--Klein modes, to achieve a 4D effective action governing the light- and heavy-light-flavor meson states. It provides an accurate prescription for the decay constants and the mass spectrum of the light-flavor, as well as for the heavy-light-flavor mesons with a charm quark constituent. Also, to describe heavy-heavy-flavor systems, Ref. \cite{Chen:2021wzj} proposed the introduction of additional fields, such as the scalar $\mathsf{H}$ field in addition to the dilaton, being responsible for differentiating the $\rho$ and the $J/\psi$ meson states.  Once the DCE of the $K_1$ and $f_1$ meson families is computed for each radial quantum number $n$, one can interpolate the obtained values and compose what is called in the literature as the  DCE-Regge trajectories of first type \cite{Bernardini:2018uuy}. Also expressing the DCE as a function of the experimental mass spectra of the $K_1$ and $f_1$ meson families, one can again  interpolate the  data points based on the range of the discrete set of the known DCE, therefore constituting DCE-Regge trajectories of second type. 
The strategy for obtaining the mass spectrum of the next generation of meson resonances in the $K_1$ and $f_1$ families consists of extrapolating the two types of Regge trajectories for higher values of $n$.  
Extrapolating the DCE-Regge trajectories yields the mass spectrum of the resonances with $n>3$, in the strange axial-vector kaon $K_1$ family, when the interpolation formul\ae\; of the DCE-Regge trajectories of first and second types are worked out for the states $K_1(1270), K_1(1400),$ and $K_1(1650)$ reported in the PDG \cite{pdg}. Regarding the $f_1$ meson family, values $n>5$ must be considered, since  there are five resonances $f_1(1285)$, $f_1(1420)$, $f_1(1510)$, $f_1(1970)$, and  $f_1(2310)$ in the PDG \cite{pdg}.

This paper is organized as follows: Sec. \ref{sec1}   introduces the main ingredients of the bottom-up 4-flavor AdS/QCD.  The mass spectrum of strange axial-vector kaon meson resonances, $K_1$, and the $f_1$ meson family as well are acquired by the solution of a coupled system of Schr\"odinger-like equations,  which are read off the equation of motions (EOMs) governing the $K_1$ and $f_1$ mesons.  
 Sec. \ref{sec2} briefly addresses the DCE apparatus. The DCE for both the strange axial-vector kaon family, $K_1$, and the $f_1$ meson families are  then evaluated first as a function of $n$, yielding DCE-Regge trajectories of first type. Thereafter, the DCE of the $K_1$ and the $f_1$ meson families are described as a function of their respective mass spectra, whose interpolation method originates DCE-Regge trajectories of second type, for each meson family. The mass spectrum of the next generation of mesonic resonances in the $K_1$ and the $f_1$ meson families are then obtained when the DCE-Regge trajectories are extrapolated for values of the radial quantum number that are higher than the ones labeling the experimentally measured resonances, in both the $K_1$ and the $f_1$ meson families, in the PDG. 
 Sec. \ref{iv} addresses final remarks and conclusions.

\section{4-flavor AdS/QCD}
\label{sec1}
In Refs. \cite{Chen:2021wzj,Ballon-Bayona:2017bwk,Momeni:2020bmy}  a holographic QCD model for four flavors $N_f=4$ was established, where the ground states and the higher excitation states of light-flavor mesons, light-heavy-flavor mesons, and the charmonium can be obtained.
Given a hard wall cutoff $z_m$, the background is defined by the 5D action \cite{Chen:2021wzj}
	\begin{eqnarray}\label{action}
		S \!&\!=\!&\! -\int_\upepsilon^{z_m}\!\!\! d^5 x \sqrt{-g}  e^{-\phi}\, {\rm Tr}\, \left[ |D\mathsf{X}|^{2}  \!-\! 3 |\mathsf{X}|^2 \!+\!|D\mathsf{H}|^2 \!-\! 3 |\mathsf{H}|^2\right.\nonumber\\&&\left.\qquad \qquad\qquad+\frac{1}{4 g_5^2} \left(\mathsf{L}^{MN} \mathsf{L}_{MN} + R^{MN} \mathsf{R}_{MN} \right)
		\right],
	\end{eqnarray}
with  coupling constant $g_5^2=12\pi^2/N_c$ \cite{Erlich:2005qh}. 
The covariant derivatives of $\mathsf{X}$ and $\mathsf{H}$ are given by  $D_{M} \mathsf{X}=\partial_M \mathsf{X} -i\mathsf{L}_M \mathsf{X} +i \mathsf{X} \mathsf{R}_M$, $D_M \mathsf{H} = \partial_M \mathsf{H} - i {V_M^{15}} \mathsf{H} - i \mathsf{H} {V_M^{15}}$, 	whereas the Yang--Mills field strengths read 
$\mathsf{L}_{MN} = \partial_{[M} \mathsf{L}_{N]} - i \left [\mathsf{L}_M , \mathsf{L}_N \right]$, $\mathsf{R}_{MN} = \partial_{[M} \mathsf{R}_{N]}- i \left [ \mathsf{R}_M , \mathsf{R}_N \right ]$,  for $\mathsf{L}_M=\mathsf{L}_M^at^a$ and $\mathsf{R}_M=\mathsf{R}_M^at^a$, being $t^a$ (\cat{$a=1,\ldots,15$}) the generators of the $SU(4)$ group. In terms of the vector and the axial-vector fields, one can still express $\mathsf{R}_M = V_M-A_M$ and $\mathsf{L}_M = V_M + A_M$.  \cat{The scalar field $\mathsf{H}$ is introduced in the action (\ref{action}) to designate the imbalance among the light- and heavy-flavor quark masses. Since we work in the 4-flavor AdS/QCD setup, namely $N_f=4$, the scalar field $\mathsf{X}$ breaks the $SU(4)_L\times SU(4)_R\simeq SU(4)_V\times SU(4)_A$ symmetry of the system to $SU(4)_V$. The residual symmetry yields the EOMs describing the light-flavor vector mesonic states to  be identical to the EOMs modeling the charmonium. For this reason, the $\mathsf{H}$ scalar field must be taken into account to distinguish the $\rho$ from the  $J/\psi$ mesons \cite{Chen:2021wzj}. Equivalently, the difference between light- and heavy-flavor quark masses is thus implemented by the distance between the light- and heavy-flavor branes in the $D_4$-$D_8$ Sakai--Sugimoto  model \cite{Liu:2016iqo}. Here the field $\mathsf{H}$ plays the role of the $\Psi$ field, in the Dirac--Born--Infeld action.  
In an equivalent manner, the $\mathsf{H}$ field is responsible for breaking the residual $SU_V(4)$ symmetry to $SU_V(3)$.}  The gauge
invariance in the 5D theory translates into the current conservation under a global symmetry, in the dual 4D theory.

Following Ref. \cite{Erlich:2005qh}, the scalar field $\mathsf{X}$ promotes a mapping to the operator $\langle\bar{q}_\mathsf{R}q_\mathsf{L}\rangle$ on the boundary, while the two gauge fields $\mathsf{L}_M^a$ and $\mathsf{R}_M^a$ correspond to $\langle\bar{q}_{\mathsf{L},\mathsf{R}}\gamma_\mu t^aq_{\mathsf{L},\mathsf{R}}\rangle$ operator. The nonzero vacuum expectation value of the field $\mathsf{X}$ describes the chiral symmetry breaking of light-flavor quarks. The novelty in the 4-flavor AdS/QCD comprises both a quadratic dilaton, $\phi(z)=\mu^2z^2$, as first proposed by Ref. \cite{Karch:2006pv}, and the hard wall cutoff $z_m$ \cite{Polchinski:2001tt}, to be concomitantly considered, yielding the possibility of  simultaneously obtaining the mass spectra of light- and heavy-flavor meson resonances. In this context, the 4-flavor AdS/QCD emulates part of both the hard and the soft wall apparatus. The field $\mathsf{X}$ can be written as
	\begin{eqnarray}\label{eqkk}
		\mathsf{X} = e^{i\uppi^bt^b} \, \mathsf{X}_0 \, e^{i\uppi^at^a},
	\end{eqnarray}
where $\mathsf{X}_0={\rm diag}[\upsilon_l(z),\upsilon_l(z),\upsilon_s(z),\upsilon_c(z)]$ and 
\beq\label{eqk}
\lim_{z\to0}\upsilon_{l,s,c}(z)=M_{l,s,c}\,z+\Upsigma_{l,s,c}\,z^3,
\eeq at the UV boundary limit.  The indexes $l,s,c$ respectively denote the light-flavor $d$ and $u$, the strange, and the  charm quarks, composing the 4-flavor AdS/QCD. Eq. (\ref{eqk}) comes from the expectation value of the field $\mathsf{X}$ in Eq. (\ref{eqkk}), dictated by the classical solution satisfying the UV boundary condition $2\mathsf{X}(\upepsilon) = M\upepsilon$, for the quark mass matrix, $M$, given by $\mathsf{X}_0(z) = \frac12 Mz + \frac12 \Upsigma z^3$, where $\Upsigma^{\rho\sigma} = \langle \bar{q}^\rho q^\sigma\rangle$ \cite{Karch:2006pv}. Therefore $\mathsf{H}={\rm diag}[0,0,0,h_c(z)]$, where $\lim_{z\to 0}h_c(z)=m_c z$ \cite{Chen:2021wzj}. 

The Poincar\'e patch metric endows a slice of the  AdS${}_5$ bulk spacetime, given by
	\begin{eqnarray}\label{eq.1}
		ds^2=\frac{L^2}{z^2}\left(dz^2+\upeta_{\mu\nu}dx^{\mu}\,dx^{\nu}\right),
	\end{eqnarray}
with $\mu,\nu=0,1,2,3$,  for $\upeta_{\mu\nu}=\rm{diag}[-1,1,1,1]$. Hereon  $L=1$ will be considered, without loss of generality. Note that this model does  not consider the backreaction of the dilaton field on the metric. Certainly, a step forward in the direction of a more realistic model would be considering an actual solution of the Einstein-dilaton gravity. Ref. \cite{Li:2011hp} was the first work to figure out the gravity background with the warp factor $z^2$ in holographic QCD, effectively implementing and solving the Einstein-dilaton system. Besides, other studies approaches in this context were implemented in Refs. \cite{dePaula:2008fp, Gursoy:2007cb, He:2013qq,Li:2013oda,Zollner:2018uep,Mahapatra:2018gig,dePaula:2010yu,Ballon-Bayona:2023zal}. Following Refs. \cite{Gherghetta:2009ac, Ballon-Bayona:2023zal},  one can implement chiral symmetry breaking and still have a quadratic dilaton in the IR, which has a pivotal role in describing the available light-flavor meson spectra.
In the 4-flavor AdS/QCD, the fields of the model can be described in terms of the meson fields as \cite{Ballon-Bayona:2017bwk,Chen:2021wzj}
	\begin{widetext}
	\begin{eqnarray}
		V &=&  \mathsf{V}^a t^a  = \frac{1}{\sqrt 2}
		\left ( \begin{matrix}
			\frac{{\scalebox{.97}{${\scalebox{.97}{$\rho^0$}}$}}}{\sqrt{2}} + \frac{{\scalebox{.97}{$\omega'$}}}{\sqrt{6}} + \frac{{\scalebox{.97}{$\psi$}}}{\sqrt{12}}  &  {\scalebox{.97}{$\rho^{_{+}}$}}  &  {\scalebox{.97}{$K^{_{*+}}$}}  &  {\scalebox{.97}{$\bar D^{_{*0}}$}}  \\
			\rho^{_{-}}   & -\frac{{\scalebox{.97}{$\rho^0$}}}{\sqrt{2}}  + \frac{{\scalebox{.97}{$\omega'$}}}{\sqrt{6}}  + \frac{{\scalebox{.97}{$\psi$}}}{\sqrt{12}}   &  {\scalebox{.97}{$K^{_{*0}}$}}  &  D^{_{*-}}  \\
			{\scalebox{.97}{$K^{_{*-}}$}}  &  {\scalebox{.97}{$\bar{K}^{_{*0}}$}}  & - \sqrt{\frac23} {\scalebox{.97}{$\omega'$}} + \frac{{\scalebox{.97}{$\psi$}}}{\sqrt{12}}  &  {\scalebox{.97}{$D_s^{_{*-}}$}}  \\
			{\scalebox{.97}{$D^{_{*0}} $}}&  {\scalebox{.97}{$D^{_{*+}} $}}  &  {\scalebox{.97}{$D_s^{_{*+}}  $}} & - \frac{3}{\sqrt{12}} {\scalebox{.97}{$\psi$}}
		\end{matrix} \right ), 	\end{eqnarray}
\end{widetext}	
\begin{widetext}
	\begin{eqnarray}
		A &=&  {\mathsf{A}}^a t^a  =  \frac{1}{\sqrt 2} \left(\begin{array}{cccc}
			\frac{ {\scalebox{.97}{$a^0_1$}}}{\sqrt{2}}  + \frac{{\scalebox{.97}{$f_1$}}}{\sqrt{6}}+\frac{{\scalebox{.97}{$\chi_{c1}$}}}{\sqrt{12}}
			&  {\scalebox{.97}{$a^{_{+}}_1$}} & {\scalebox{.97}{$K_{1}^{_{+}}$}}& {\scalebox{.97}{$\bar {\scalebox{.97}{${D}_{1}^{_{0}}$}} $}}\\
			{\scalebox{.97}{$a_1^{_{-}}$}}&- \frac{ {\scalebox{.97}{$a^{_{0}}_1$}}}{\sqrt{2}} +\frac{{\scalebox{.97}{$f_1$}}}{\sqrt{6}}+\frac{{\scalebox{.97}{$\chi_{c1}$}}}{\sqrt{12}}
			& {\scalebox{.97}{$K_{1}^{_{0}} $}}& {\scalebox{.97}{$D_{1}^{_{-}}$}}\\
			{\scalebox{.97}{$K_{1}^{_{-}}$}}&{\scalebox{.97}{$ \bar{K}_{1}^{_{0}}$}}  & -\sqrt{\frac23}{\scalebox{.97}{$f_1$}}+\frac{{\scalebox{.97}{$\chi_{c1}$}}}{\sqrt{12}}& {\scalebox{.97}{$D_{s1}^{_{-}}$}}\\
			{\scalebox{.97}{${D}_{1}^{_{0}}$}}&{\scalebox{.97}{$D_{1}^{_{+}}$}}&{\scalebox{.97}{$D_{s1}^{_{+}}$}}&- \frac{3}{\sqrt{12}}{\scalebox{.97}{$\chi_{c1}$}}
		\end{array}
		\right), \\
		\uppi &=& \uppi^a t^a = \frac{1}{\sqrt 2}
		\left (\begin{matrix}
			\frac{{\scalebox{.97}{$\pi^{{0}}$}}}{\sqrt{2}}  + \frac{{\scalebox{.97}{$\eta$}}}{\sqrt{6}}  + \frac{{\scalebox{.97}{$\eta$}}_c}{\sqrt{12}}  & {\scalebox{.97}{$ \pi^{_{+}}$}}  &  {\scalebox{.97}{$K^{_{+}} $}}  &  \bar {\scalebox{.97}{$D^{_{0}}$}}  \\
			{\scalebox{.97}{$\pi^{_{-}}$}}   & -\frac{{\scalebox{.97}{$\pi^0$}}}{\sqrt{2}}  + \frac{{\scalebox{.97}{$\eta$}}}{\sqrt{6}}  + \frac{{\scalebox{.97}{$\eta$}}_c}{\sqrt{12}}   &  {\scalebox{.97}{$K^{_{0}} $}} & {\scalebox{.97}{$ D^{_{-}} $}} \\
			{\scalebox{.97}{$K^{_{-}}$}}  &  {\scalebox{.97}{$\bar K^{_{0}}$}}  & - \sqrt{\frac23} {\scalebox{.97}{$\eta$}} + \frac{{\scalebox{.97}{$\eta$}}_c}{\sqrt{12}}  &  {\scalebox{.97}{$D_s^{_{-}} $}} \\
			{\scalebox{.97}{$D^{_{0}}$}} &  {\scalebox{.97}{$D^{_{+}}$}}  &  {\scalebox{.97}{$D_s^{_{+}}$}}  & - \frac{3}{\sqrt{12}} {\scalebox{.97}{$\eta$}}_c
		\end{matrix} \right ).
	\end{eqnarray}
\end{widetext}
	The mesons mass spectra as well as the 3- and 4-point coupling constants are obtained by expanding the action (\ref{action}) to 4${}^{\rm th}$ order (see Ref. \cite{Chen:2021wzj} for details). For the scalar vacuum expectation value $\upsilon_{l,s,c}$, the EOMs read 
	\begin{eqnarray}
	\left[\partial_z^2 +e^{-\phi(z)}\left(\frac{3}{z^4}-\frac{\phi'(z)}{z^3}\right)\partial_z
	-\frac{m_5^2}{z^2}\right]\upsilon_q(z)=0,
	\end{eqnarray}
whose solutions are given in terms of Tricomi's (confluent hypergeometric) function, $U$, and the associated Laguerre polynomial, $L$, as
	\begin{eqnarray}
		\upsilon_q(z)=c_1(q)\sqrt{\pi}z U\left(\frac12,0,\phi\right)-c_2(q)z L\left(-\frac12,-1,\phi\right) \, .
	\end{eqnarray}
		Denoting $M_A^{ab}={\rm Tr}\left([t^a,\mathsf{X}_0][t^b, \mathsf{X}_0]\right)$, the EOMs of the transverse part of axial-vector fields are\begin{widetext}
	\begin{eqnarray}\label{eoma}
		\left[\partial_z^2 -e^{-\phi(z)}\left(\frac{1}{z^2}-\frac{\phi'(z)}{z}\right)\partial_z
+\frac{2g_5^2M_A^{ab}}{z^2}\right] {\mathsf{A}}^a_{\mu{\scalebox{.55}{$\perp$}}}(z,q)=-q^2{\mathsf{A}}^a_{\mu{\scalebox{.55}{$\perp$}}}(z,q),
	\end{eqnarray}\end{widetext}
where ${\mathsf{A}}^{z,a} = 0$ and $\partial_\mu \mathsf{A}_{\scalebox{.55}{$\perp$}}^{\mu,a}=0$ (gauge choice), denoting the Fourier transform by ${\mathsf{A}}^a_{\mu{\scalebox{.55}{$\perp$}}}(z,x)=\frac{1}{(2\pi)^4}\int d^4q e^{iqx}{\mathsf{A}}^a_{\mu{\scalebox{.55}{$\perp$}}}(z,q)$. 
	The fields ${\mathsf{A}}^a_{\mu{\scalebox{.55}{$\perp$}}}(z,q)$ can be split off as ${\mathsf{A}}^a_{\mu{\scalebox{.55}{$\perp$}}}(z,q)={\mathsf{A}}^{(\textsc{n})a}_{\mu{\scalebox{.55}{$\perp$}}}(z,q)+{\mathsf{A}}^{0a}_{\mu{\scalebox{.55}{$\perp$}}}{\mathsf{A}}^a_{{\scalebox{.55}{$\perp$}}}(z,q)$, where the index $\textsc{n}$ here labels the Kaluza--Klein excitations.  The Dirichlet and Neumann boundary conditions are respectively given by $\lim_{z\to0}{\mathsf{A}}^{(\textsc{n})a}_{\mu{\scalebox{.55}{$\perp$}}}(z,q)=0$ and $\lim_{z\to z_m}\partial_z{\mathsf{A}}^{(\textsc{n})a}_{\mu{\scalebox{.55}{$\perp$}}}(z,q)= 0$. 
	
	The EOMs of the longitudinal part of the axial-vector fields and the pseudoscalar fields are, respectively, written as:
	\begin{widetext}
	\begin{subequations}
	\begin{eqnarray}	
		q^2z^2\partial_z \upphi^a(z,q)+{2g_5^2M_A^{ab}}\partial_z \uppi^a(z,q)&=&0\,,\label{eom-pi-1}\\			
		\left[z^2\partial_z^2 -e^{-\phi(z)}\left({1}-z{\phi'(z)}\right)\partial_z\right]\upphi^a(z,q) +
		{2g_5^2M_A^{ab}}\left(\uppi^a(z,q)-\upphi^a(z,q)\right)&=&0\,,\label{eom-pi-2} 					
	\end{eqnarray}
	\end{subequations}
	\end{widetext}
	where the spacetime derivative $\partial_\mu\upphi^a$ of the field $\upphi^a$ is the longitudinal component of the axial-vector field $A_\mu^a$.   The boundary conditions  $\lim_{z\to0}\uppi^{(\textsc{n})a}(z,q)=\lim_{z\to0}\upphi^{(\textsc{n})a}(z,q)=0$ and $\lim_{z\to z_m}\partial_z\upphi^{(\textsc{n})a}(z,q)= 0$ are considered, for $\textsc{n}\neq 0$. For the Goldstone mode, the latter condition is replaced by $\lim_{z\to z_m}\partial_z\upphi^{(\textsc{n})a}(z,q)\propto \lim_{z\to z_m}\partial_z\uppi^{(\textsc{n})a}(z,q)$.  
	
	By fixing the parameters of the model, the masses of the pseudoscalar, the vector, and the axial-vector mesons, as well as their excited states with higher values of $n$, can be obtained. For it, the related physical parameters comprise the quadratic dilaton coefficient $\mu$, the infrared hard wall cutoff $z_m$, and the vacuum expectation values $c_1(q)$ for $q=(l, s, c)$. These parameters are fixed by the experimental masses of the ${\scalebox{.97}{$\psi$}}(3770)$, $J/{\scalebox{.97}{$\psi$}}(1S)$, ${\scalebox{.97}{$\chi_{c1}$}}(3872)$, $K^{*}(892)$, $\rho(770)$, and $a_{1}(1260)$ resonances. The value  $\mu\approxeq0.43$ GeV can be chosen for both the mass and the slope of the Regge trajectory of the $\rho$ meson to be fitted. 	Also, the hard wall cutoff, $z_m$ can be acquired by the  $J/{\scalebox{.97}{$\psi$}}$ mass and the slope of its Regge trajectory, together with the ${\scalebox{.97}{$\psi$}}(3770)$ meson state. By Eq. (\ref{eqk}), the values $c_1(l)$, $c_1(s)$, and $c_1(c)$ can be read off the quark masses $M_{l,s,c}$, as well as  the value of $\Upsigma_{l,s,c}$, encoding the quark condensation. For it, the values adopted are $M_{l}=0.140~$GeV, $M_{s}=0.2~$GeV, $M_{c}=1.2~$GeV, whereas $\Upsigma_{l}= 2.4603\times 10^6\rm{MeV}^{3}$, $\Upsigma_{s}=3.5118\times 10^6~\rm{MeV}^{3}$,  $\Upsigma_{c}=2.1024\times 10^7~\rm{MeV}^{3}$,   $m_c=1.020~$GeV and $\sigma_{c}=1.7984\times 10^7~\rm{MeV}^{3}$  \cite{Chen:2021wzj}.
	The model predictions and experimental data of the mesons masses are listed in Tables \ref{scalarmasses1} and \ref{scalarmasses2}, respectively for the $K_1$ and $f_1$ meson families in PDG \cite{pdg}.

\begin{table}[H]
\begin{center}\medbreak
\begin{tabular}{||c|c||c|c|c||}
\hline\hline
$n$ & State & $M_{\scalebox{.67}{\textsc{Exp.}}}$ (MeV)  & $M_{\scalebox{.67}{\textsc{AdS/QCD}}}$ (MeV) &\;RE (\%)\;\\
       \hline\hline
\hline
1 &\;$K_1(1270)\;$ & $1272\pm7$ & 1328  & 4.2 \\ \hline
2 &\;$K_1(1400)\;$ & $1403\pm7 $ & 1557 & 9.9 \\ \hline
3& \;$\;K_1(1650)$& $1672 \pm 50$       & 1763 &  5.1   \\\hline
\hline\hline
\end{tabular}
\caption{Mass spectrum of strange axial-vector kaons resonances: experimental values and 4-flavor AdS/QCD prediction.  The fifth column displays the relative error. } \label{scalarmasses1}
\end{center}
\end{table}
It is worth pointing out that the mass of the $K_1(1270)\;$ state is the average calculated by the PDG. Depending on the form of production, the value of the mass can differ, as the $K_1(1270)$ is produced either by $K$ beams (and by beams other than $K$ mesons) or by lepton decays \cite{pdg}. 

\begin{figure}[H]
	\centering
	\includegraphics[width=7.5cm,height=5.8cm]{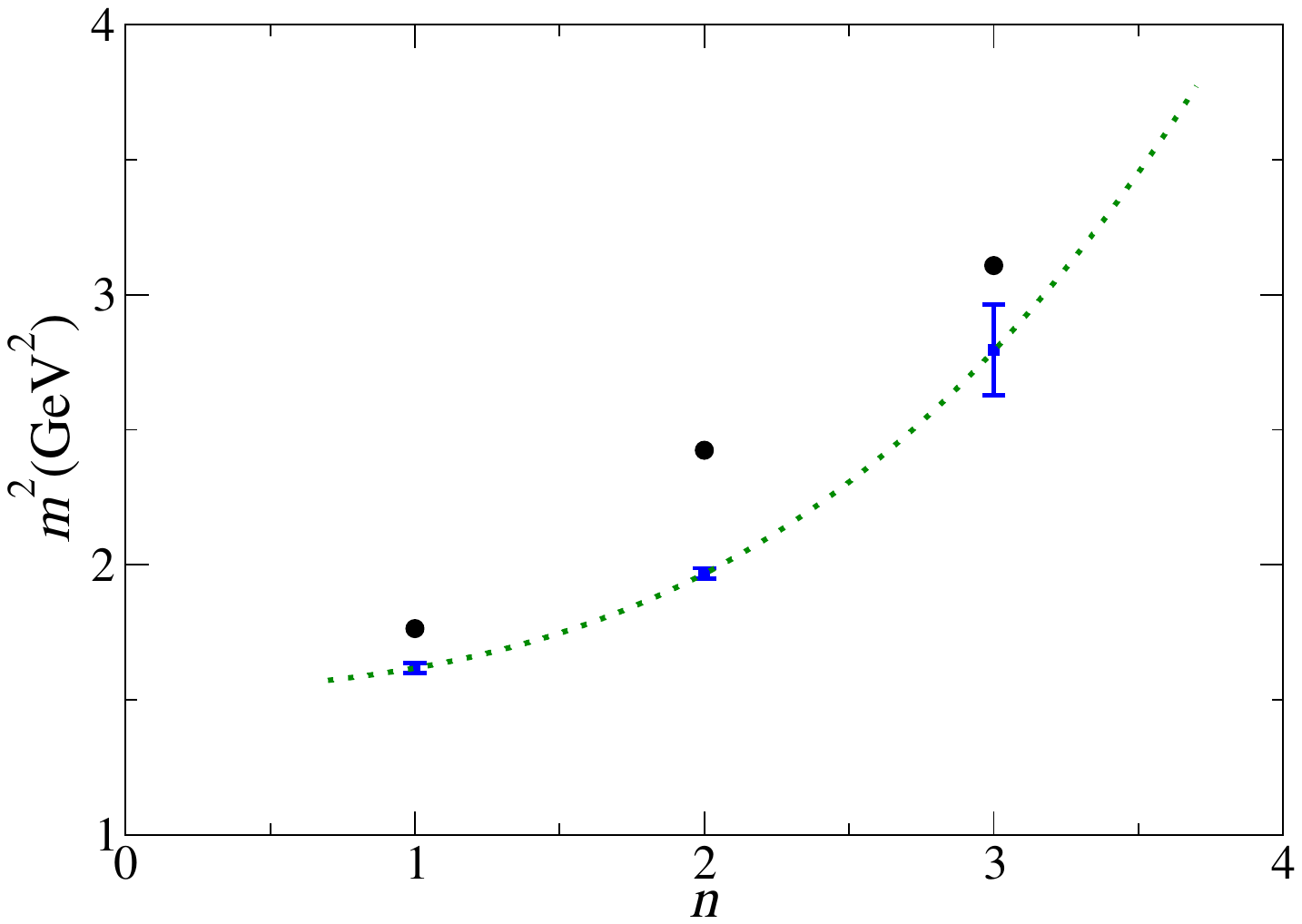}
	\caption{Mass spectrum of strange axial-vector kaons radial resonances ($K_1$ meson family). Squares: experimental values with error bars. Circles: 4-flavor AdS/QCD prediction for $n=1,2,3$. Dotted line: interpolation function given by Eq. (\ref{lrt}).}
	\label{cen1}
\end{figure}

The  nonlinear Regge trajectory is illustrated in Fig. \ref{cen1}, together with the mass spectrum of strange  vector kaon resonances predicted by the 4-flavor AdS/QCD.
The Regge trajectory, relating the squared mass spectrum of the $K_1$ meson family as a function of $n$, fits experimental data in PDG \cite{pdg}, and reads  
\beq \label{747}
m_n^2 = 0.0371 \, n^3 + 0.0152 \, n^2 + 0.0443 \, n + 1.521,\label{lrt}
\eeq
within a 0.1\% root-mean-square deviation (RMSD). 

\begin{figure}[H]
	\centering
	\includegraphics[width=7.5cm,height=5.8cm]{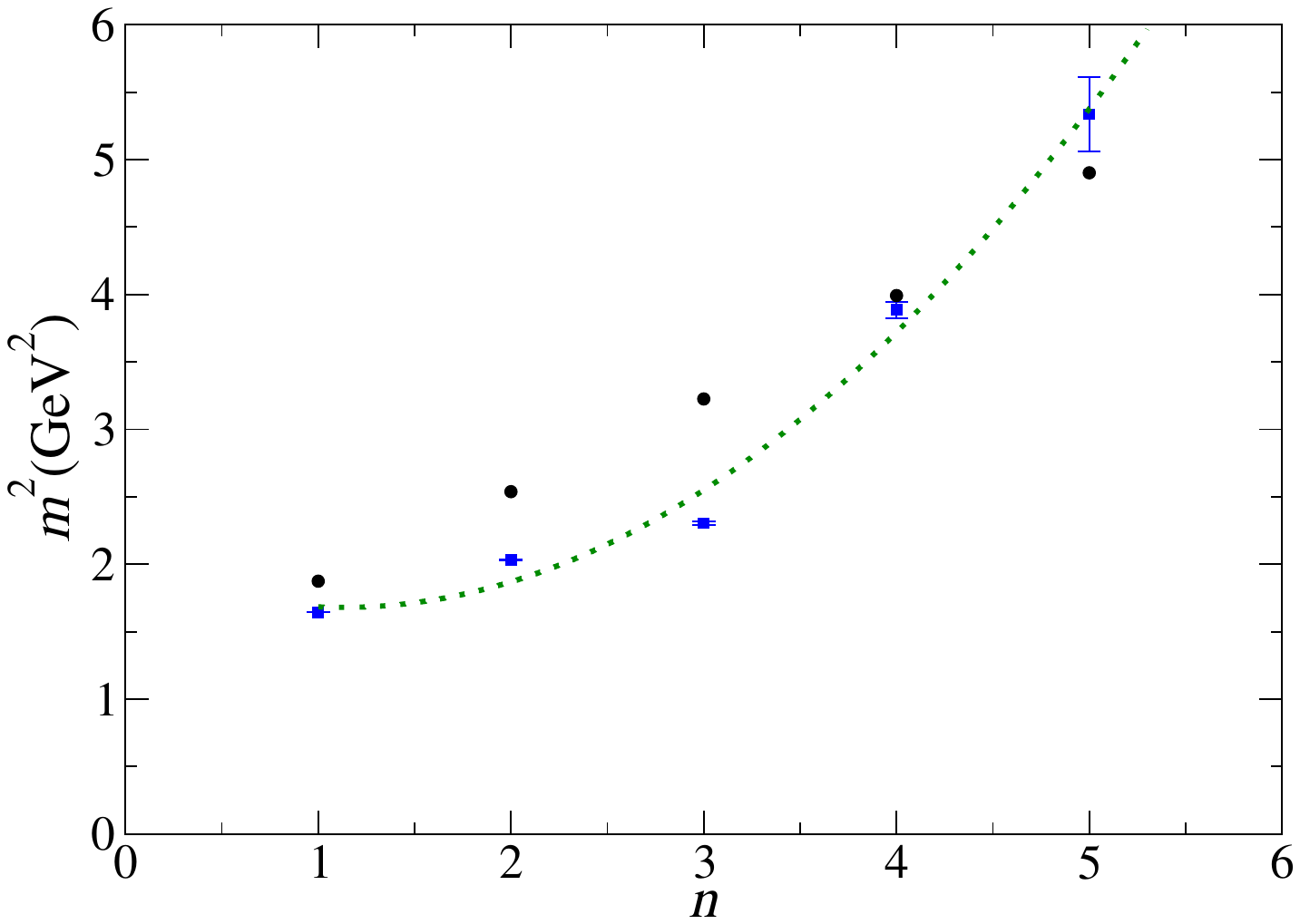}
	\caption{Mass spectrum of $f_1$ meson resonances.  Squares: experimental values with error bars. Circles: 4-flavor AdS/QCD prediction for $n=1,\ldots,5$. Dotted line: interpolation function given by Eq. (\ref{lrt1}).}
	\label{cen2}
\end{figure}

The $f_1$ meson family can be now addressed.
\begin{table}[H]
\begin{center}\medbreak
\begin{tabular}{||c|c||c|c|c||}
\hline\hline
$n$ & State & $M_{\scalebox{.67}{\textsc{Exp.}}}$ (MeV)  & $M_{\scalebox{.67}{\textsc{AdS/QCD}}}$ (MeV) &\;RE (\%)\;\\
       \hline\hline
\hline
1 &\;$f_1(1285)\;$ & $1281.9\pm0.5$ & 1369  & 6.3 \\ \hline
2 &\;$f_1(1420)\;$ & $1426.3\pm0.9 $ & 1593 & 10.5 \\ \hline
3& \;$\;f_1(1510)$& $1518 \pm 5$       & 1796 &  15.5   \\\hline
4& \;$\;f_1(1970)$& $1971 \pm 15$       & 1998 &  1.3   \\\hline
5& \;$\;f_1(2310)$& $2310 \pm 60$       & 2214 &  4.1   \\\hline
\hline\hline
\end{tabular}
\caption{Mass spectrum of the $f_1$ meson family: experimental values and 4-flavor AdS/QCD prediction.  The fifth column displays the relative error. } \label{scalarmasses2}
\end{center}
\end{table}
\noindent The last two resonances, $f_1(1970)$ and $f_1(2310)$, are omitted from the Summary Table in PDG. The Regge trajectory, now relating the squared mass spectrum of the $f_1$ meson family to $n$, is illustrated in Fig. \ref{cen2}, together with the mass spectrum of $f_1$ meson resonances predicted by the 4-flavor AdS/QCD model.
The predictions due to the 4-flavor AdS/QCD yield the  nonlinear Regge trajectory, which  relates the squared mass of $f_1$ as a function of $n$, given by   
\beq \label{747}
m_n^2 &=& -6.821\times 10^{-4} \, n^3 + 0.2512 \, n^2\nonumber\\&& - 0.5629 \, n + 1.996,\label{lrt1}
\eeq
within a 5.9\% root-mean-square deviation (RMSD).

\section{DCE of $K_1$ and $f_1$ meson resonances}
\label{sec2}
DCE-based techniques can be employed to determine the mass spectrum of the next generation of meson resonances in the $K_1$ and $f_1$ families. To achieve it, the energy density describing the meson resonances in the 4-flavor AdS/QCD is taken into account and, subsequently, its Fourier transform must be calculated. The DCE paradigm involves describing the physical system in terms of its wave modes. If one denotes by ${\scalebox{.95}{$\boldsymbol{q}$}}$ the spatial part of the 4-momentum $q^\mu$, the Fourier transform is, as usual, described as  
\beq\label{fou}
{\scalebox{.92}{$\tau_{00}$}}({{\scalebox{.95}{$\boldsymbol{q}$}}}) = \frac{1}{(2\pi)^{\ell/2}}\int_{\scalebox{.72}{${\mathbb{R}^\ell}$}}\,{\scalebox{.92}{$\tau_{00}$}}({\boldsymbol{r}})e^{-i{{\scalebox{.65}{$\boldsymbol{q}$}}}\cdot{\scalebox{.65}{${\boldsymbol{r}}$}}}\,d^\ell{\scalebox{.95}{${\boldsymbol{r}}$}}.\eeq 
The so-called  modal fraction is the next step to compute the DCE and is related to the power spectral density given by    
\cite{Gleiser:2012tu,Gleiser:2011di,Gleiser:2018kbq}, 
\begin{eqnarray}
{\scalebox{.91}{${\boldsymbol{\tau}}_{00}$}}({\boldsymbol{{q}}}) = \frac{\left|{\scalebox{.92}{$\tau_{00}$}}({{\scalebox{.95}{$\boldsymbol{q}$}}})\right|^{2}}{ \bigintsss_{\scalebox{.72}{${\mathbb{R}^\ell}$}}  \left|{\scalebox{.92}{$\tau_{00}$}}({\bf{q}})\right|^{2}\,d^\ell{{\scalebox{.935}{${\bf q}$}}}}.\label{modalf}
\end{eqnarray} 
As a consequence, the DCE can evaluate the amount of information that is demanded for encoding the energy density, which describes  the physical system under scrutiny. The DCE, in this way, characterizes the weight of information entropy transported by each wave mode in the momentum space. The DCE can be  computed by the formula: 
\begin{eqnarray}
{\rm DCE}= - \int_{\scalebox{.72}{${\mathbb{R}^\ell}$}}\,{\scalebox{.91}{${\check{\boldsymbol{\tau}}}_{00}$}}({{\scalebox{.935}{${\bf q}$}}})\log   {\scalebox{.91}{${\check{\boldsymbol{\tau}}}_{00}$}}({{\scalebox{.935}{${\bf q}$}}})\,d^\ell{{\scalebox{.935}{${\bf q}$}}},
\label{confige}
\end{eqnarray}
where {${\scalebox{.91}{$\check{{\boldsymbol{\tau}}}_{00}$}}({{\scalebox{.95}{$\boldsymbol{q}$}}})={\scalebox{.91}{${\boldsymbol{\tau}}_{00}$}}({{\scalebox{.95}{$\boldsymbol{q}$}}})/{\scalebox{.93}{${\boldsymbol{\tau}}^{\scalebox{.6}{\textsc{max}}}_{00}$}}({{\scalebox{.95}{$\boldsymbol{q}$}}})$} is the modal fraction associated to the differential configurational complexity, and ${\scalebox{.93}{${\boldsymbol{\tau}}^{\scalebox{.6}{\textsc{max}}}_{00}$}}({{\scalebox{.95}{$\boldsymbol{q}$}}})$ denotes the maximum value of the energy density 
{{\scalebox{.91}{${\boldsymbol{\tau}}_{00}$}} in ${\mathbb{R}^\ell}$. The DCE is measured in units of nat/unit volume, where \emph{nat} denotes the  natural unit  of information entropy, with 1 bit $\approxeq 0.693147$ nat \cite{Gleiser:2018kbq}.

 The value $\ell=1$ will be regarded hereon to estimate the DCE of the $K_1$ and $f_1$ meson families, according to the steps given by the subsequent use of  Eqs. (\ref{fou}) -- (\ref{confige}), due to the codimension-1 AdS boundary of the 4-flavor AdS/QCD. The  Lagrangian underlying Eq. (\ref{action}) can be replaced in the  expression of the energy density,
\beq
{\scalebox{.92}{$\tau_{00}$}}\!&=&\!  \frac{2}{\sqrt{ -g }}\!\! \left[\frac{{\scalebox{.923}{$\partial$}} (\sqrt{-g}{{\scalebox{.923}{$\mathcal{L}$}}})}{{\scalebox{.923}{$\partial$}}{g^{00}}} \!-\!\frac{{\scalebox{.923}{$\partial$}}}{{\scalebox{.923}{$\partial$}}{ x^\upalpha }}  \frac{{\scalebox{.923}{$\partial$}} (\sqrt{-g} {{\scalebox{.923}{$\mathcal{L}$}}})}{{\scalebox{.923}{$\partial$}}\left(\frac{{\scalebox{.79}{$\,{\scalebox{.923}{$\partial$}}$}} g^{00}}{{\scalebox{.79}{$\,{\scalebox{.923}{$\partial$}}$}}x^\upalpha}\right)}
  \right],
  \label{em1}
\eeq
whose Fourier transform can be evaluated by Eq. (\ref{fou}). Therefore the modal fraction and the DCE are computed when employing Eqs. (\ref{modalf}, \ref{confige}), respectively.  
 This routine allows for assessing  the DCE  underlying the 4-flavor AdS/QCD. With it in hand, the mass spectrum of the next generation of mesons in both the $K_1$ and $f_1$ families can be then obtained, by interpolating the experimental mass spectrum of the \svkps in PDG \cite{pdg}.  The DCE, using the protocol (\ref{fou}) -- (\ref{confige}) is numerically calculated. 
 We will first approach the \svkps  in the next subsection, for which the first fundamental strange axial-vector kaon state is the  $K_1(1270)$.

 \subsection{DCE of strange axial-vector kaons $K_1$}
 \label{ndce}
 We start by calculating the DCE of the $K_1$ meson family. The first {} \svka resonance, $n=1$,  corresponds to the 
$K_1(1270)$ meson. Taking $n=2$, the state $K_1(1400){}$ is described, while  $n=3$ represents the state $K_1(1650)$.  
The DCE  can be then evaluated, using the energy density \eqref{em1} into Eqs. (\ref{fou}) -- (\ref{confige}). The DCE hybridizing the 4-flavor AdS/QCD comprises a realistic scenario that may surpass bottom-up AdS/QCD estimates for the mass spectrum of the next generation of meson resonances, taking into account the experimental values of the mass spectrum of the $K_1(1270)$, $K_1(1400){}$, and $K_1(1650)$ to obtain the mass spectrum of the next generation meson resonances in the $K_1$ family. The values of the DCE, numerically computed, for the $K_1$ meson family, are listed in Table \ref{scalarmasses50}.
\begin{table}[H]
\begin{center}
\begin{tabular}{||c|c|c||}
\hline\hline
$n$ & \;{} Strange axial-vector kaons\; & DCE (nat) \\
       \hline\hline
\hline
\;1\; &\;$K_1(1270){}\;$ & 154.134   \\ \hline
\;2\; &\;$K_1(1400){}\;$ & 228.981   \\ \hline
\;3\;& \;$K_1(1650){}$& 437.755          \\\hline
\hline
\end{tabular}
\caption{DCE of the $K_1$ meson family, for $n=1,2,3$.} \label{scalarmasses50}
\end{center}
\end{table}
The DCE-Regge trajectory of first type corresponds to expressing the DCE with respect to the $n$ quantum number. Quadratic polynomial  interpolation of data in Table \ref{scalarmasses50} yields the  DCE-Regge trajectory of first type, 
\begin{eqnarray}\label{itp1}
\!\!\!\!\!\!\!\!\!\!\!\!\clt{{\rm DCE}}_{K_1}(n)&=& 66.59 \, n^2 - 124.92 \, n + 212.46,  \end{eqnarray}
 \clt{within $0.01\%$ RMSD}. Data in Table \ref{scalarmasses50} and Eq. (\ref{itp1}) are together depicted in Fig. \ref{cen1d}.

\begin{figure}[H]
	\centering
	\includegraphics[width=7.5cm,height=5.8cm]{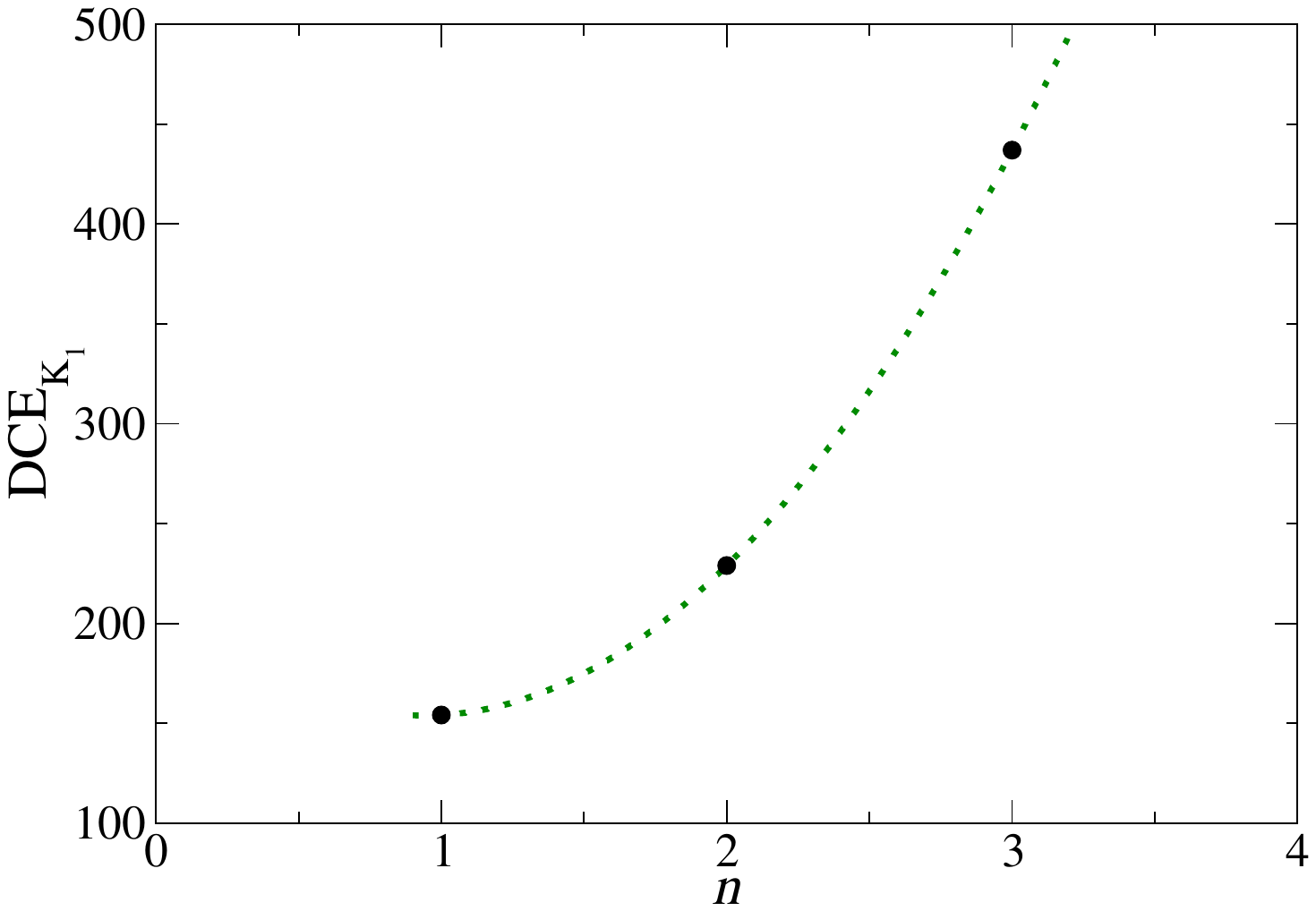}
	\caption{DCE of $K_1$ strange axial-vector kaon family as a function of  $n$, for  $n=1,2,3$ (respectively corresponding to the states $K_1(1270){}$, $K_1(1400){}$, and $K_1(1650){}$ \cite{pdg}).  
The DCE-Regge trajectory of first type, Eq. (\ref{itp1}), is plotted by a dotted line.}
	\label{cen1d}
\end{figure}
In addition, the DCE underlying the $K_1$ meson family can be arranged with respect to the experimental mass spectrum of the $K_1$ meson states. Therefore the DCE of the $K_1$ meson family, listed in Table \ref{scalarmasses50},  can be plotted, instead, as a function of the squared mass of each meson state, already accessible in Table \ref{scalarmasses1}. The resulting data are displayed in Fig. \ref{cem11}, with an interpolation curve corresponding to a DCE-Regge trajectory of second type, given by Eq. (\ref{itq11}). 
\begin{figure}[H]
	\centering
	\includegraphics[width=7.5cm,height=5.8cm]{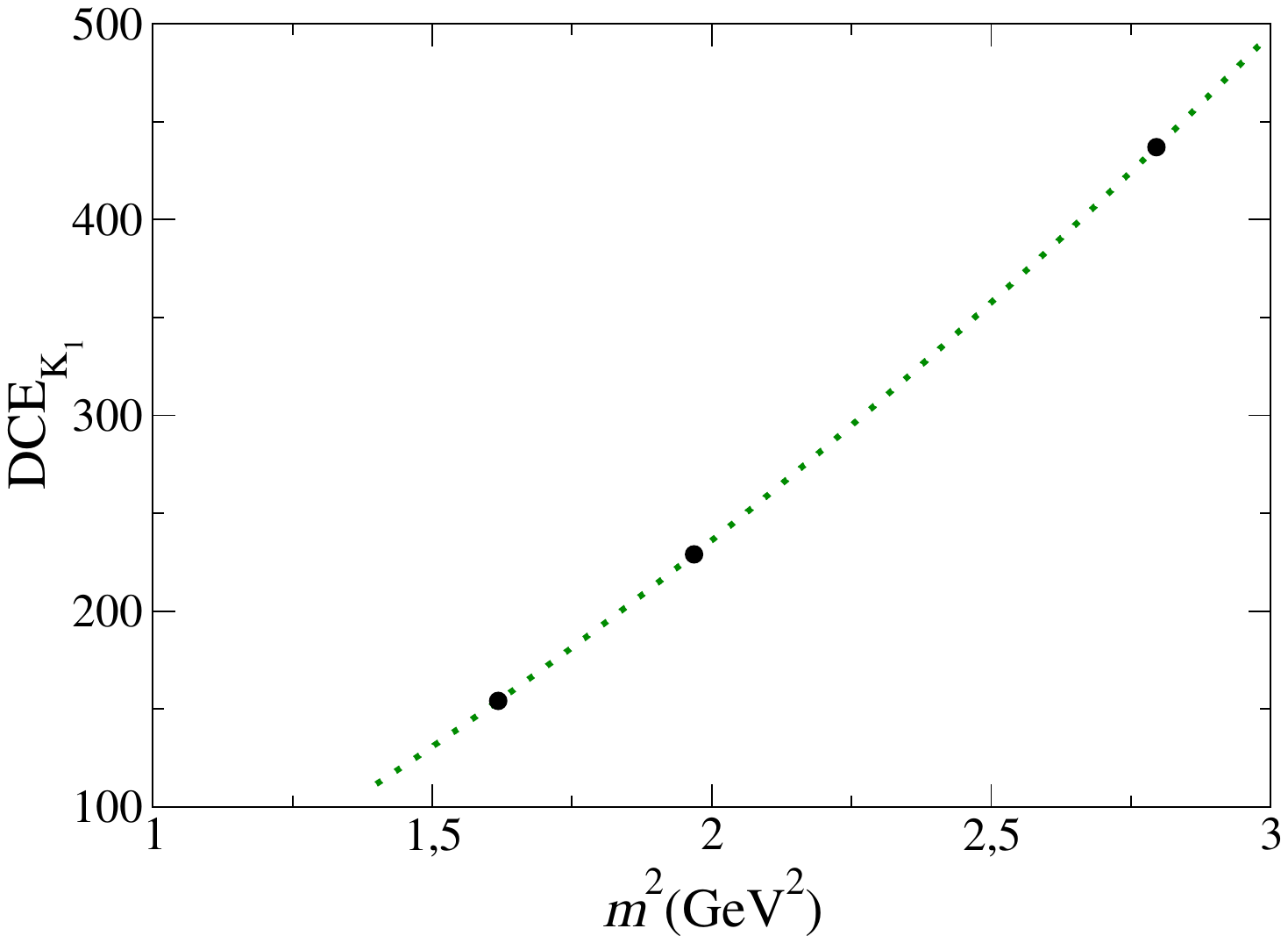}
	\caption{DCE of the $K_1$ meson family displayed as a function of their respective squared mass, for  $n=1,2,3$ (corresponding, respectively, to the $K_1(1270){}$, $K_1(1400){}$, and $K_1(1650){}$ states \cite{pdg}). 
The DCE-Regge trajectory of second type in Eq. (\ref{itq11}) corresponds to the interpolating dotted green line.}
	\label{cem11}
\end{figure}
\noindent The DCE-Regge trajectory of second type, expressing the DCE of $K_1$ meson family with respect to their respective (squared) mass spectrum, $m^2$ (GeV${}^2$), is given by \begin{eqnarray}
\label{itq11}
\!\!\!\!\!\!\!\!\!\!\!{\rm DCE}_{K_1}(m) \!\!&\!=\!&\!\!-0.585 m^6 \!+\! 35.91 m^4 \!+\! 90.45 m^2 \!-\! 83.77
   \end{eqnarray} within $0.01\%$  RMSD.    
   
Eqs. (\ref{itp1}) and (\ref{itq11}) are two types of DCE-Regge trajectories, encoding all the features necessary to deploy the mass spectrum of meson resonances in the $K_1$ family with $n>3$. When Eq. (\ref{itp1}) is taken into account, one can take values $n>3$ and read off the respective values of the DCE, replacing their respective value onto the left-hand side of Eq. (\ref{itq11}) and algebraically solve it, generating a technique for obtaining the mass of each resonance in the $K_1$ family $n>3$. This protocol is established just on the DCE, which represents the configurational entropy underlying the mesons in the 4-flavor AdS/QCD, and on the experimental mass spectrum of the first three states of the $K_1$ meson family in PDG \cite{pdg}. Therefore it is  a more realistic method when compared to the 4-flavor AdS/QCD, where the mass spectrum of $K_1$ resonances can be read off the Schr\"odinger-like equation's eigenvalues. 
Nevertheless one must emphasize that Eq.  (\ref{itq11}), representing the DCE-Regge trajectory of second type which expresses the DCE of the $K_1$ meson family as a function of their respective (squared) mass spectrum, is acquired by interpolating the experimental mass spectrum of the $K_1$ meson states, also shown in Fig. \ref{cem11}.

To obtain the  mass spectrum of the $K_1$ meson resonances for $n>3$, 
let us start with $n=4$, corresponding to the $K_1^\star$ strange axial-vector kaon resonance. Substituting $n=4$ into Eq. (\ref{itp1}), one can verify that the DCE underlying the $K_1^\star$ equals 688.221 nat. Thus, when this value is replaced in the left side of Eq. (\ref{itq11}), one can pick up the solution of the 3${}^{\rm rd}$-degree polynomial equation in the $m^2$ variable, immediately reading the value  $m_{K_1^\star}= 1903.95$ MeV for the mass of  the $K_1^\star$ state. When Eq. (\ref{itp1}) is evaluated for $n=5$, the corresponding DCE can be calculated, supplying the value 1072.910 nat. Then Eq. (\ref{itq11}) can be worked out for this value, giving the mass $m_{K_1^{\star\star}}= 2267.12$ MeV, for the $K_1^{\star\star}$  strange axial-vector kaon resonance.  Finally, Eq. (\ref{itp1}) can be estimated for $n=6$, hence providing the DCE of the $K_1^{\star\star\star}$ resonances to be equal to 1560.189 nat. Substituting it into Eq. (\ref{itq11}) results the mass  value $m_{K_1^{\star\star\star}}= 2419.97$ MeV. Resonances with $n>7$ have high values of DCE and are configurationally very  unstable. Also due to the high values of the configurational complexity, the $K_1$ meson resonances for $n>7$  are very unlikely to be produced, being highly unstable. Hence the focus here is to obtain the mass spectrum of the first three resonances of the next generation of mesonic states in the $K_1$ family. These outcomes are listed in Table 
\ref{scalarmasses102}. 
	\begin{table}[H]
\begin{center}\begin{tabular}{||c|c|c|c||}
\hline\hline
$n$ & State & $M_{\scalebox{.67}{\textsc{Exp.}}}$ (MeV)  & $M_{\scalebox{.67}{\textsc{AdS/QCD and hybrid}}}$ (MeV) \\
       \hline\hline
\hline
1 &\;$K_1(1270)\;$ & $1272\pm7$ & 1328   \\ \hline
2 &\;$K_1(1400)\;$ & $1403\pm7 $ & 1557  \\ \hline
3& \;$\;K_1(1650)$& $1672 \pm 50$       & 1763   \\\hline
4& \;$K_1^\star\;$& ---------  & 1903.95${}^\diamond$  \\\hline
5& \;$K_1^{\star\star}\;$& ---------     & 2267.12${}^\diamond$   \\\hline
6& \;$K_1^{\star\star\star}\;$&  ---------   & 2419.97${}^\diamond$ \\\hline
\hline\hline
\end{tabular}
\caption{Table \ref{scalarmasses1} added with the resonances  of the $K_1$ meson family for $n=4, 5, 6$. In the 4${}^{\rm th}$ column, the states a `` ${}^\diamond$ '' stand for the mass spectrum extrapolated by employing both Eqs.  (\ref{itp1}, \ref{itq11}), which interpolate the experimental mass spectrum of $K_1$ meson states for $n=1, 2, 3$. } \label{scalarmasses102}
\end{center}
\end{table}
The $K_1^{\star\star}$ strange axial-vector kaon resonance, whose mass has been estimated to equal 2267.12 MeV, might match both the  $X(2210)$ states in the \textsc{Other Mesons -- Further States} listing in PDG \cite{pdg}, which respectively presents experimental masses  $2210^{+79}_{-15}$ MeV and $2207\pm22$ MeV. Moreover, the $K_1^{\star\star\star}$ strange axial-vector kaon resonance, has mass predicted to equal 2203.43  MeV. Hence it may also correspond to the  second $X(2210)$ meson resonance reported in the PDG, with experimental mass value $2210^{+79}_{-15}$ MeV \cite{pdg}.

 \subsection{DCE of $f_1$ meson resonances}
 \label{ndce}
 
Now the $f_1$ meson family will be addressed, whose mass spectrum of its next generation of resonances will be derived in the context of the DCE. 
The first state, $n=1$, in this meson family  corresponds to the  $f_1(1285)$ meson, which typically fits into the standard quark model as a constituent of the $^3P_1$ (axial-vector) nonet. It also represents the isoscalar flavor mixing equivalent of the state of the $f_1$ meson family with $n=2$, namely, the $f_1(1420)$.The $f_1(1285)$ state was found in the process of $p\bar{p}$ annihilation, with a given resonance which decays into $K\bar{K}\pi$, identified to  $I^G(J^{PC}) = 0^+(1^{++})$ \cite{CLAS:2016zjy}.  More recent experiments reported measurement of the $f_1(1285)$, in the context of $pp$ production and photon collisions as well \cite{pdg}.  One can evaluate the DCE of the $f_1$ meson resonances already detected and reported in the PDG \cite{pdg}, for  $n=1,\ldots,5$,  respectively representing the states 
$f_1(1285)$, $f_1(1420)$, $f_1(1510)$, $f_1(1970)$, and  $f_1(2310)$, where the last two resonances are in the \textsc{Other Mesons -- Further States} listing in PDG \cite{pdg}. 

Using  Eqs. (\ref{fou}) -- (\ref{confige}), the DCE underlying  the $f_1$ meson family is numerically computed and listed in Table \ref{scalarmasses501}.
\begin{table}[H]
\begin{center}
\begin{tabular}{||c|c|c||}
\hline\hline
$n$ & \;{} $f_1$ meson family\; & DCE (nat) \\
       \hline\hline
\hline
1 &\;$f_1(1285)\;$ &  47.124 \\ \hline
2 &\;$f_1(1420)\;$ & 52.109 \\ \hline
3& \;$\;f_1(1510)$&  57.465  \\\hline
4& \;$\;f_1(1970)$& 85.921  \\\hline
5& \;$\;f_1(2310)$& 122.578  \\\hline
\hline
\end{tabular}
\caption{DCE of the states in the $f_1$ meson family with $n=1,\ldots,5$.} \label{scalarmasses501}
\end{center}
\end{table}
The  DCE-Regge trajectory of first type, this time related to $f_1$ mesons,  displays the DCE as a function dependent on $n$. Fig. \ref{cen1dd} arrays corresponding results, whose polynomial  interpolation of $3{}^{\rm rd}$-order, taking into account data in Table \ref{scalarmasses501}, comprises the DCE-Regge  trajectory of first type, given by  
\begin{eqnarray}\label{itp11}
\!\!\!\!\!\!\!\!\!\!\!\!\!\!\!\!\clt{{\rm DCE}}_{f_1}(n)\!&\!=\!&\! 0.6508 n^3 \!+\! 0.3167 n^2 \!-\! 3.214 n \!+\! 49.90,  \end{eqnarray}
 \clt{within $0.02\%$ RMSD}.

\begin{figure}[H]
	\centering
	\includegraphics[width=7.5cm,height=5.8cm]{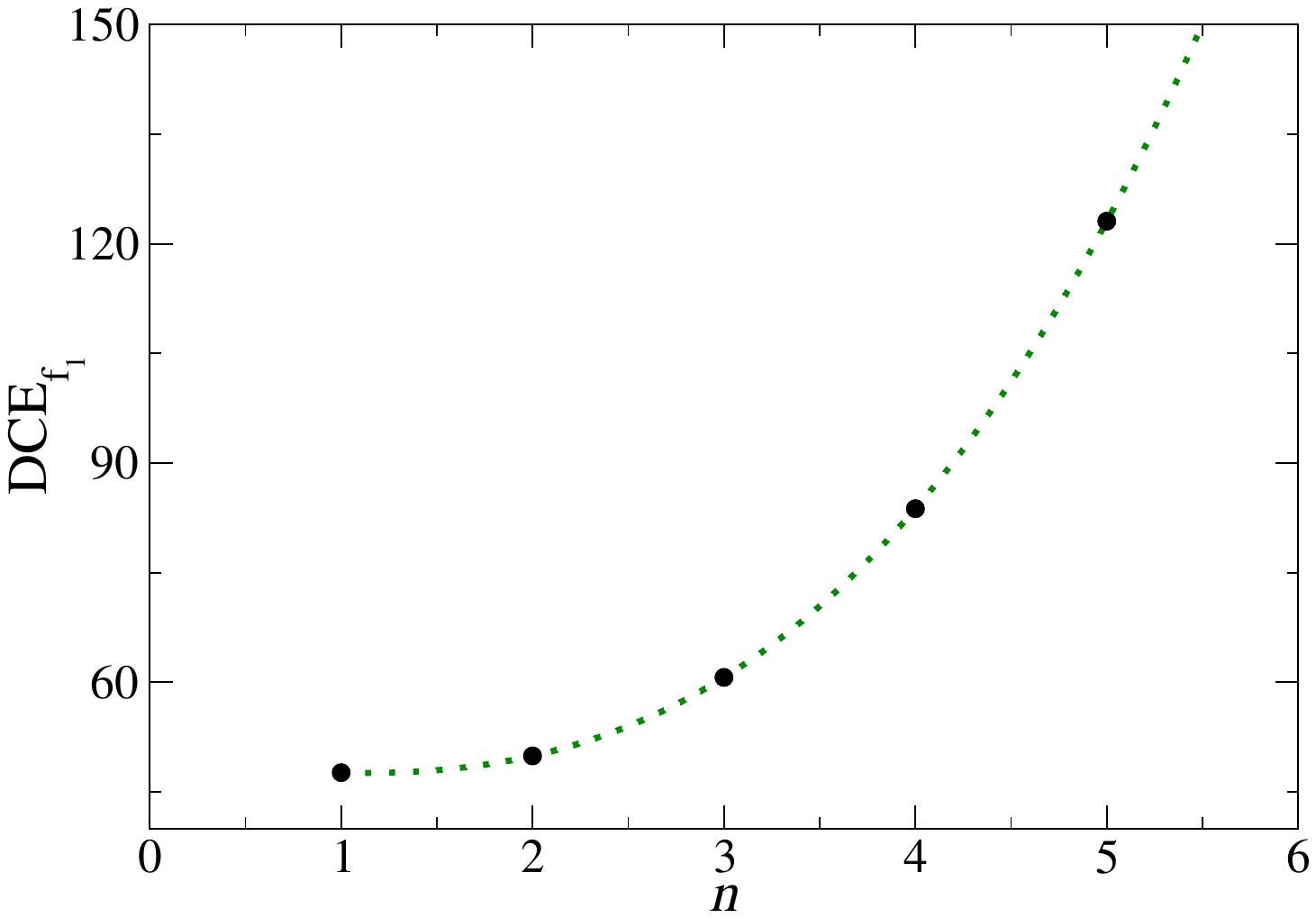}
	\caption{DCE of the $f_1$ meson family, for  $n=1,\ldots,5$ (respectively corresponding to the 
$f_1(1285)$, $f_1(1420)$, $f_1(1510)$, $f_1(1970)$, and  $f_1(2310)$ states  in PDG \cite{pdg}).  
The DCE-Regge trajectory of first type (\ref{itp11}) is displayed as the interpolating  dotted green  line.}
	\label{cen1dd}
\end{figure}
Furthermore, the DCE underlying the $f_1$ meson states can be understood as a function of the experimental mass spectrum of these meson states, accounting for the DCE-Regge trajectory of second type. Taking into account the DCE of the set of states in the $f_1$ meson family in Table \ref{scalarmasses501}, it can be also realized as a mass-dependent function for each $f_1$ meson resonance, immediately available through Table \ref{scalarmasses2}. The outcome is presented in Fig. \ref{cem12}. Interpolating the discrete data yields the DCE-Regge trajectory of second type, with the interpolation formula given in Eq. (\ref{itq112}). 
\begin{figure}[H]
	\centering
	\includegraphics[width=7.5cm,height=5.8cm]{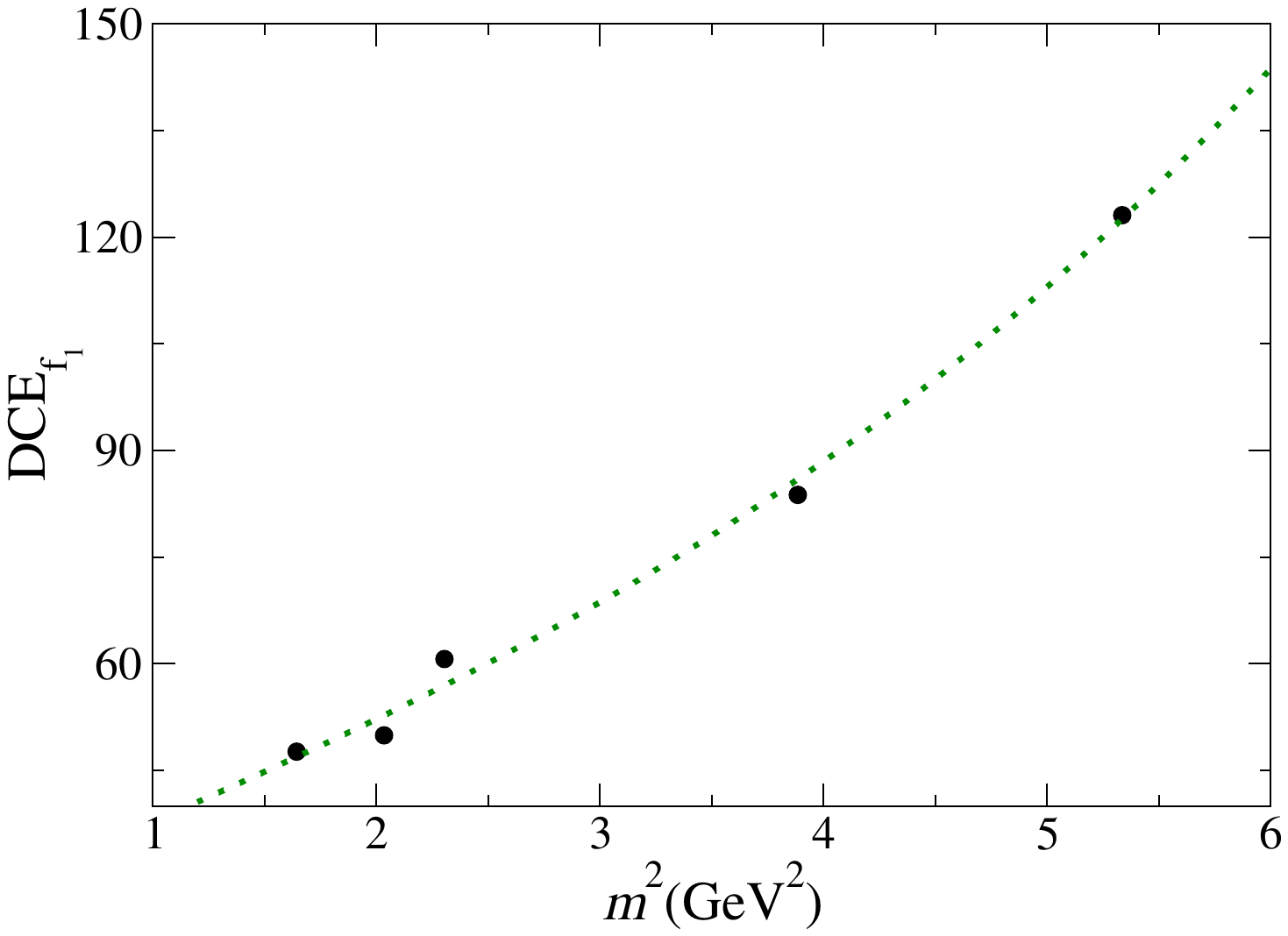}
	\caption{DCE of the $f_1$ meson states as a function of their squared mass, for  $n=1,\ldots,5$ (respectively corresponding to the 
$f_1(1285)$, $f_1(1420)$, $f_1(1510)$, $f_1(1970)$, and  $f_1(2310)$ states,  in PDG \cite{pdg}). 
The DCE-Regge trajectory of second type, in Eq. (\ref{itq112}), corresponds to the interpolating dotted green line.}
	\label{cem12}
\end{figure}
\noindent The DCE-Regge trajectory of second type, which makes the DCE of the $f_1$ meson family to be dependent on the squared mass spectrum of these states, $m^2$ (GeV${}^2$), is given by 
\begin{eqnarray}
\label{itq112}
\!\!\!\!\!\!\!\!\!\!\!\!{\rm DCE}_{f_1}(m) \!&\!=\!&0.232 m^3 \!-\! 0.407 m^2 \!+\! 14.06 m \!+\! 23.88, 
   \end{eqnarray} within $1.8\%$  RMSD.    
By the use of Eq. (\ref{itp11}), the DCE of the $f_1$ meson resonances  can be inferred and these values, for each $n>5$, can be superseded into the left-hand side of the DCE-Regge trajectory of second type (\ref{itq112}), which can be solved. This DCE-based technique yields the mass spectrum of $f_1$ meson resonances for  $n>5$, relying upon the experimental mass spectrum of the $f_1$ meson resonances. To calculate the value of the mass of the  $f_1^\star$ resonance, corresponding to putting $n=6$ in Eq. (\ref{itp11}) yields the DCE equals 182.60  nat. Hence, this value in Eq. (\ref{itq112}) gives the mass value $m_{f_1^\star}= 2648.84$ MeV. For finding the mass of the $f_1^{\star\star}$ resonance, Eq. (\ref{itp11}) must be evaluated with $n=7$, originating the DCE value 266.16 nat. Therefore Eq. (\ref{itq112}) can be worked out, yielding the mass $m_{f_1^{\star\star}}= 2944.02$ MeV. Finally, Eq. (\ref{itp11}) the $n=8$ resonance can be considered in Eq. (\ref{itp11}), giving the DCE of the $f_1^{\star\star\star}$ to assume 377.69 nat. When Eq. (\ref{itq112}) is resolved for this value of the DCE, the mass $m_{f_1^{\star\star\star}}= 3206.94$ MeV is read off. Resonances with $n>9$ present very high values of DCE, being very unstable from the configurational point of view. Due to the complexity of shape of these resonances for $n>9$, they are very unlikely to be produced, due to their very high instability. Therefore we focus on the mass spectrum of the first three resonances of the next generation of mesonic states in the $f_1$ family. All these results are listed in Table 
\ref{scalarmasses1022}. 
	\begin{table}[H]
\begin{center}\begin{tabular}{||c|c|c|c||}
\hline\hline
$n$ & State & $M_{\scalebox{.67}{\textsc{Exp.}}}$ (MeV)  & $M_{\scalebox{.67}{\textsc{AdS/QCD and hybrid}}}$ (MeV) \\
       \hline\hline
\hline
1 &\;$f_1(1285)\;$ & $1281.9\pm0.5$ & 1369   \\ \hline
2 &\;$f_1(1420)\;$ & $1426.3\pm0.9 $ & 1593 \\ \hline
3& \;$\;f_1(1510)$& $1518 \pm 5$       & 1796    \\\hline
4& \;$\;f_1(1970)$& $1971 \pm 15$       & 1998    \\\hline
5& \;$\;f_1(2310)$& $2310 \pm 60$       & 2214    \\\hline
6& \;$f_1^\star\;$& ---------  & 2648.84${}^\diamond$  \\\hline
7& \;$f_1^{\star\star}\;$& ---------     & 2944.02${}^\diamond$   \\\hline
8& \;$f_1^{\star\star\star}\;$&  ---------   & 3206.94${}^\diamond$ \\\hline
\hline\hline
\end{tabular}
\caption{Table \ref{scalarmasses2} completed with $f_1$ resonances with $n>5$.  The  masses extrapolated from the DCE-Regge trajectories (\ref{itp11}, \ref{itq112}), for $n=6,7,8$, in the 4${}^{\rm th}$ column, are indicated with a `` ${}^\diamond$ ''. } \label{scalarmasses1022}
\end{center}
\end{table}
Extrapolating the DCE-Regge trajectories (\ref{itp11}, \ref{itq112}) for $n=6$ has yielded the mass of the $f_1^{\star}$ meson resonance to be equal to 2248.84 MeV. This state might emulate the  $X(2680)$ meson state in the \textsc{Other Mesons -- Further States} listing, in PDG \cite{pdg}, presenting mass  $2676\pm27$ MeV, experimentally measured. Taking into account the RMSD in Eq. (\ref{itp11}), another feasible possibility is the $X(2632)$ state, which presents mass equal to   $2635.2\pm3.3$ MeV and may be identified to the $f_1^{\star}$ state, according to the PDG. Lastly, the $f_1^{\star\star\star}$ resonance, with $n=8$, has mass 3206.94  MeV obtained using DCE-techniques in the 4-flavor AdS/QCD and may  match the  $X(3250)$ state in PDG \cite{pdg}, with mass  $3240\pm8\pm20$ MeV experimentally measured and reported in PDG \cite{pdg}.

\section{Conclusions}\label{iv}

The DCE used in the context of the 4-flavor AdS/QCD took into account the experimental mass spectrum of the strange axial-vector mesons $K_1(1270), K_1(1400),$ and $K_1(1650)$, and of the family $f_1(1285)$, $f_1(1420)$, $f_1(1510)$, $f_1(1970)$, and  $f_1(2310)$,
 to predict the mass spectrum of resonances with higher radial quantum numbers in both these meson families. This approach, besides having more precision than solving the EOMs to obtain the mass spectrum, is phenomenologically  robust, since it relies on the experimental mass spectrum of both families in PDG \cite{pdg}.  Some possibilities of identifying the obtained next generation of meson resonances to further meson states, in listings of the PDG,  were addressed and discussed. 
  This hybrid technique, taking into account the DCE of the meson resonances, computed from the energy-momentum tensor regulating the 4-flavor AdS/QCD, was implemented by the concomitant analysis  of DCE-Regge  trajectories of first and second type, extrapolating them for $n=6,7,8$, from the respective interpolation curves for the $f_1$ family, whereas  $n=4,5,6$ was 
  implemented for the $K_1$ family. Also, DCE-based techniques are more concise, if one compares them to the obtention of the mass spectra of different families of meson resonances in the pure 4-flavor AdS/QCD paradigm, which in general takes into account the eigenvalues of the Schr\"odinger-like equations, which come from the EOMs that governs the fields. The  DCE protocol, in addition, has the advantage of taking into account more phenomenological aspects, as it takes into account the mass spectrum of the $K_1$ and $f_1$ meson families, experimentally measured and reported in the PDG. An important improvement to the 4-flavor AdS/QCD model may be implemented by taking into account the backreaction of the dilaton onto the bulk geometry, by solving the Einstein-dilaton equations, emulating the relevant results in Ref. \cite{Li:2011hp}. Then, one could use the DCE analyses for the energy-momentum tensor of this new dynamical model.

\medbreak
\paragraph*{Acknowledgments} 
GK thanks FAPESP  (grant No. 2018/19943-6) and UFABC, for the hospitality. WdP acknowledges the partial support of CNPq (Grant No. 313030/2021-9) and  CAPES (Grant No. 88881.309870/2018-01). RdR~is grateful to  FAPESP (Grant No. 2021/01089-1 and No. 2022/01734-7) and CNPq (Grant No. 303390/2019-0), for partial financial support. 
\bibliography{bibliography}

\end{document}